\documentstyle[11pt,paspconf,epsfig]{article}

\begin{document}

\title{Gamma-Ray Burst Lines}

\author{Michael S. Briggs}
\affil{Department of Physics, University of Alabama in Huntsville,
Huntsville, AL 35899}

\begin{abstract}
The evidence for spectral features in gamma-ray bursts is summarized.
As a guide for evaluating the evidence,
the properties of gamma-ray detectors and the methods of analyzing
gamma-ray spectra are reviewed.
In the 1980's, observations indicated
that absorption features below 100 keV were present in a large 
fraction of bright gamma-ray bursts.
There were also reports of emission features around 400 keV.
During the 1990's the situation has become much less clear.
A small fraction of bursts observed with BATSE have statistically 
significant low-energy features, but the reality of the features is suspect
because in several cases the data of the BATSE detectors appear to 
be inconsistent.
Furthermore, most of the possible features appear in emission rather than
the expected absorption.
Analysis of data from other instruments has either not 
been finalized or has not detected lines.
\end{abstract}


\section{Introduction}

Formerly, lines were widely believed to exist in GRB spectra
and were regarded as
strong evidence that GRBs originate on nearby neutron stars with
intense magnetic fields.
Currently, lines are unfashionable due to a lack of recent good observations
and due to the strong evidence that most or all
GRBs originate at cosmological distances.
If the community once again becomes convinced of the existence of lines,
there would probably be a vigorous debate between a nearby neutron star
origin for a subset of GRBs and efforts to model line formation in
fireballs at cosmological distances.

I will discuss the evidence for lines in the spectra of gamma-ray bursts,
focusing on observations and analysis rather than
physical or theoretical interpretation. 
Previous reviews include
Teegarden (1982), 
Teegarden (1984),
Harding, Petrosian \& Teegarden (1986),
Higdon and Lingenfelter (1990),
Fishman and Meegan (1995) and
Briggs (1996).
While the problems involved in interpreting gamma-ray spectra and the
accepted analysis procedures should be well known
(e.g., Fenimore et al. 1983;
Teegarden 1984; Fenimore et al. 1988; Loredo \& Epstein 1989;
Briggs 1996),
I shall review these topics because they are crucial for interpreting
the evidence for GRB lines.

\section{Detector Properties}

In optical astronomy, one uses a prism or grating to spread the photons out by
wavelength---a line is simply detected as an deficit or excess in the intensity
of photons at one wavelength compared to neighboring wavelengths.
In contrast, gamma-ray detectors typically detect individual photons.
Each photon may interact in the detector by a variety of processes:
the photoelectric effect, Compton scattering, or pair production.
The energy deposited in the detector by a photon
is referred to as the ``energy loss''---the
great complication is that in some cases the energy loss may only be a fraction
of the energy of the incident photon.
A detected photon is referred to as a ``count'': we can never be sure that
the energy of a particular count equals the energy of the incident photon.
The observed counts form an energy loss spectrum from which we must 
deduce the incident photon spectrum.

As an example, I will use the Spectroscopy Detectors (SDs) of BATSE.
The principles described result from the physics of photon interactions
and are broadly applicable to all gamma-ray detectors.
The SDs are 12.7 cm diameter by 7.6 cm thick crystals of NaI(Tl) scintillator
viewed by a single photomultiplier tube.
When a particle interacts in the crystal, the resulting ionization is
converted into scintillation light and amplified by the photomultiplier.
The electronic pulse is further amplified and then digitized by the
pulse height amplifier.  The digital value is called the ``channel''
of the event. 
To first order, the scintillation light is proportional to the ionization,
so that the the channel value is nearly
proportional to energy deposited in the crystal .
Accurate analysis requires a calibration that accounts for
nonlinearities in the emission of the scintillation light
(e.g., Band et al. 1992 for the  BATSE SDs).

\begin{figure}[tb!]
\mbox{
\epsfig{file=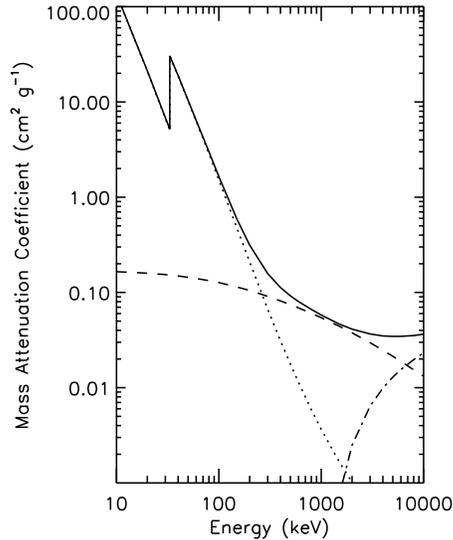,height=7.2cm,
bbllx=60bp,bblly=100bp,bburx=555bp,bbury=700bp,clip=}
\begin{minipage}[b]{79mm}
\caption{The mass attenuation coefficient $\mu$ for gamma-rays in
NaI:
solid line: total coefficient; dotted line: the photoelectric effect;
dashed line: Compton scattering;  dot-dash line: pair production.
The fractional transmission is $\exp {(-\mu x)}$, where $x$ is
the quantity of NaI transversed in g~cm$^{-2}$.  The data are from
Hubbell (1969).  Other materials will differ in detail but will also have
successive regions in which the photoelectric effect, Compton scattering
and pair production dominate.
\protect \vspace*{8mm}}
\end{minipage}
}
\end{figure}

Below about 200~keV, most photon interactions will be via
the photoelectric effect (Fig.~1).
Compton scattering is important above several hundred keV;
pair production is important only for photons with energies of
at least several MeV.

In the case of the photoelectric effect, the gamma-ray energy is totally
transfered to an atomic electron, typically an inner shell electron.
An atomic cascade will result in which fluorescent X-rays or Auger electrons
will be emitted.
The left curve of Fig.~2 illustrates a simple case, the count spectrum
expected for 20 keV photons incident on the crystal.
The line detected from a beam of monoenergetic photons is broadened
by Poisson fluctuations in the number of photoelectrons in the photomultiplier
tube.

\begin{figure}[tb!]
\mbox{
\epsfig{file=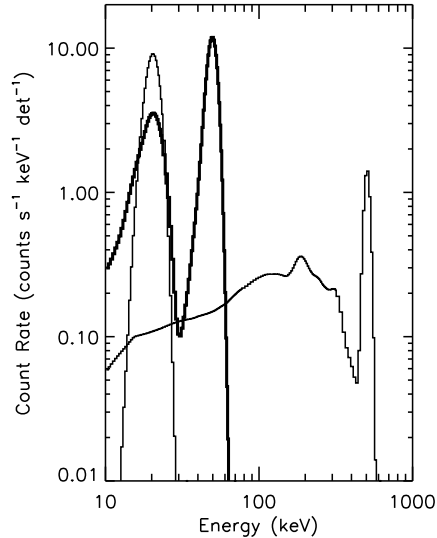,height=7.2cm,
bbllx=70bp,bblly=100bp,bburx=550bp,bbury=700bp,clip=}
\begin{minipage}[b]{79mm}
\caption{Simulated energy loss spectra for monoenergetic photons
incident on a BATSE Spectroscopy Detector.
The model response includes the photon interactions within the NaI
detector, scattered photons from the spacecraft and the Earth's atmosphere
and Poisson fluctuations in the production of photoelectrons in the
photomultiplier tube.
Left curve: incident photon energy 20 keV;
middle, bold curve: 50 keV; right curve: 500 keV.
In each case the incident flux is 1 photon s$^{-1}$ cm$^{-2}$.
\protect \vspace*{8mm}}
\end{minipage}
}
\end{figure}

Even the simple photoelectric effect can result in a complicated energy
deposition spectrum.
The 20~keV photon of the previous example will typically interact
with an electron of the L-shell of iodine---the K-shell is energetically
forbidden.
A photon with an energy greater than
the 33.17 binding energy of the iodine K-shell, such as a
50 keV photon (bold curve, Fig.~2) will typically interact with a
K-shell electron.
The resulting cascade will usually involve a L, M or N to K shell transition
and a fluorescence X-ray with an energy from 28.3 to 33.0 keV.
If this fluorescence photon also interacts in the crystal, the entire 
energy of the 50 keV may be captured (right peak of the bold curve of Fig.~2).
However, the energy fluorescence photon has a lower interaction cross-section
than the 50 keV photon (Fig.~1), and may escape the crystal, resulting
in incomplete energy deposition (left peak of the bold curve of Fig.~2).

Full energy deposition normally  occurs for photons with energies
below the 33.17 keV K-shell energy; photons with energies just above the
K-shell energy will normally interact but have a substantial probability
of incomplete energy deposition due to the escape of a fluorescence
photon.
A flat or hard incident photon spectrum will lead to a detected count spectrum
with a deficit above 33 keV.
This  feature (Fig.~3) is a detector property and should not be
mistaken for an astrophysical feature.

\begin{figure}[tb!]
\centerline{
\epsfig{file=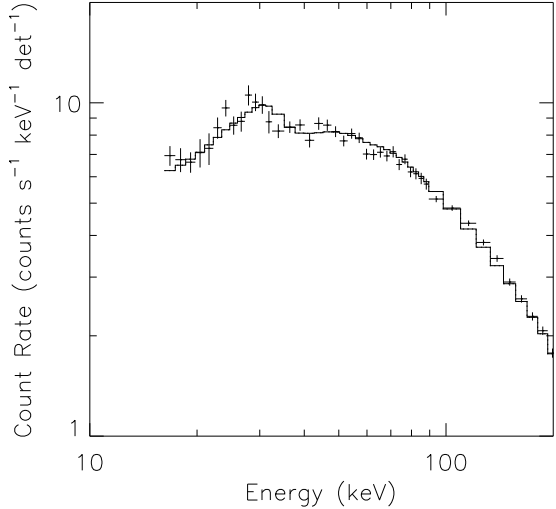,height=7cm,%
bbllx=50bp,bblly=190bp,bburx=540bp,bbury=650bp,clip=}
}
\caption{Observed and modeled energy loss spectrum for GRB~930922,
as observed with a BATSE SD.
The points depict the observed count rate, while the histogram
shows the count rate model obtained by folding a photon model through
the detector response model.
The peak at 30 keV is due to a decreased response from 35 to 50 keV
originating from the escape of fluorescence photons, as described
in the text.}
\end{figure}

At energies of a few hundred keV to a few MeV, Compton scattering is
the most likely process.
As a scattering rather than absorption process, only a portion of the
energy of the incident photon is transferred to the electron.
Fig.~2 shows the energy loss spectrum of 500 keV photons (right curve).
The peak at 500 keV originates from cases in which the energy of the
scattered photon is also captured in the detector from some combination
of Compton scattering and photoelectric interactions.
The maximum energy transfer to the scattered electron, 331 keV, occurs for the
case that the incident photon is back-scattered at $180^\circ$.
Single scattering events with angles near $180^\circ$ cause the peak
at 310 keV, while the valley above 310 keV exists because single scattering
events cannot create counts with energies above 331 keV.
The minimum energy of the scattered  photon, 169 keV, also occurs for
180$^\circ$ backscattering, so photons that interact outside the detector 
and scatter into the detector
with scattering angles near 180$^\circ$ create the peak at 190 keV.

The result of the several interaction processes is that incident photons
may deposit in the detector any energy from zero up to their entire energy.
Mapping the expected count spectra as a function of incident energy
(Fig.~4), partial energy deposition creates
off-diagonal terms in addition to the desired diagonal response.
Furthermore, Poisson fluctuations in the number of photoelectrons
produced in the photomultiplier tube cause a broadening of the
diagonal response, visible in Figures~2 and~4.
These effects are included in a computer model of a detector
by simulations using a standard
Monte Carlo particle transport code and a description
incorporating the geometry and materials of the detector
(e.g., Pendleton et al. 1995 for the BATSE detectors).

\begin{figure}[tb!]
\centerline{
\epsfig{file=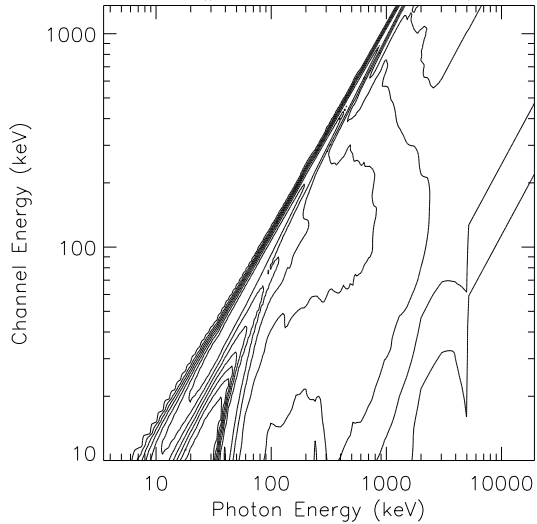,height=7cm,
bbllx=35bp,bblly=155bp,bburx=580bp,bbury=685bp,clip=}
}
\caption{A contour map of the response of a BATSE spectroscopy detector
(logarithmically spaced contours).
The units of the response are cm$^2$ keV$^{-1}$; for a fixed incident
photon energy, integrating the response with respect to the detected count
energy (channel) will yield the effective area, typically about
120 cm$^2$.
The main grouping of contours, starting at photon energy of 8 keV and
running diagonally across the figure,
is the direct or full-energy loss response.   It is broadened by
Poisson fluctuations.   The second diagonal group of contours, which
starts at photon energy of 35 keV, represents cases in which a 
fluorescence photon escapes the detector.  The remaining contours originate
from other cases of partial energy deposition in the detector.}
\end{figure}

\section{Methodology}

The lack of a one-to-one correspondence between the energy
of an incident photon and the energy of the detected count
complicates understanding the observed energy loss or count spectra.
The detected count spectrum cannot be simply inverted
into a measured photon spectrum, as it can for optical photons.
The energy of a single photon cannot be deduced from the 
energy of the observed count; 
if many counts are observed, the incident spectrum
can be unfolded subject to certain limitations.

Several direct inversion techniques have been proposed
(Loredo \& Epstein, 1989), but practitioners
have generally not found them to be useful.
Instead, the `forward-folding' deconvolution
procedure is generally used:
one {\it assumes} a photon spectral form and compares it to the data.
A model photon spectrum is represented with a parameterized function,
the model photon spectrum is converted into a model count spectrum using
a computer model of the detector, and the model and observed count
spectra are compared.
The comparison is quantified by using a statistic such as $\chi^2$ or
likelihood.   The fit is optimized, subject to using the assumed spectral
form, by varying the parameters of the function.
If a  sufficiently low value of $\chi^2$, or high value
of likelihood, is achieved, one is said to have a `good' fit.
Other spectral forms might obtain similar or better values of the fitting
statistic, so a `good' fit can never be known to be the best fit,
or to be the correct representation of the incident photon spectrum.

Sometimes the fitted photon function is used to scale the count rate data
points into what appear to be photon flux points.
These must be treated with great caution since what appear to be data
points are actually model dependent.
In particular, when lines with intrinsic widths comparable to the detector
resolution are fit, the values of the photon points scaled from the count
rate data points become strongly dependent on the model---the reality of a
spectral feature should never be judged on a plot of deconvolved
photon flux (Fig.~5).
Properly speaking, the results of the deconvolution procedure are the
parameter values of the assumed functional form and the
value of the statistic measuring the quality of the fit.

\begin{figure}[tb!]
\centerline{
\mbox{
\epsfig{file=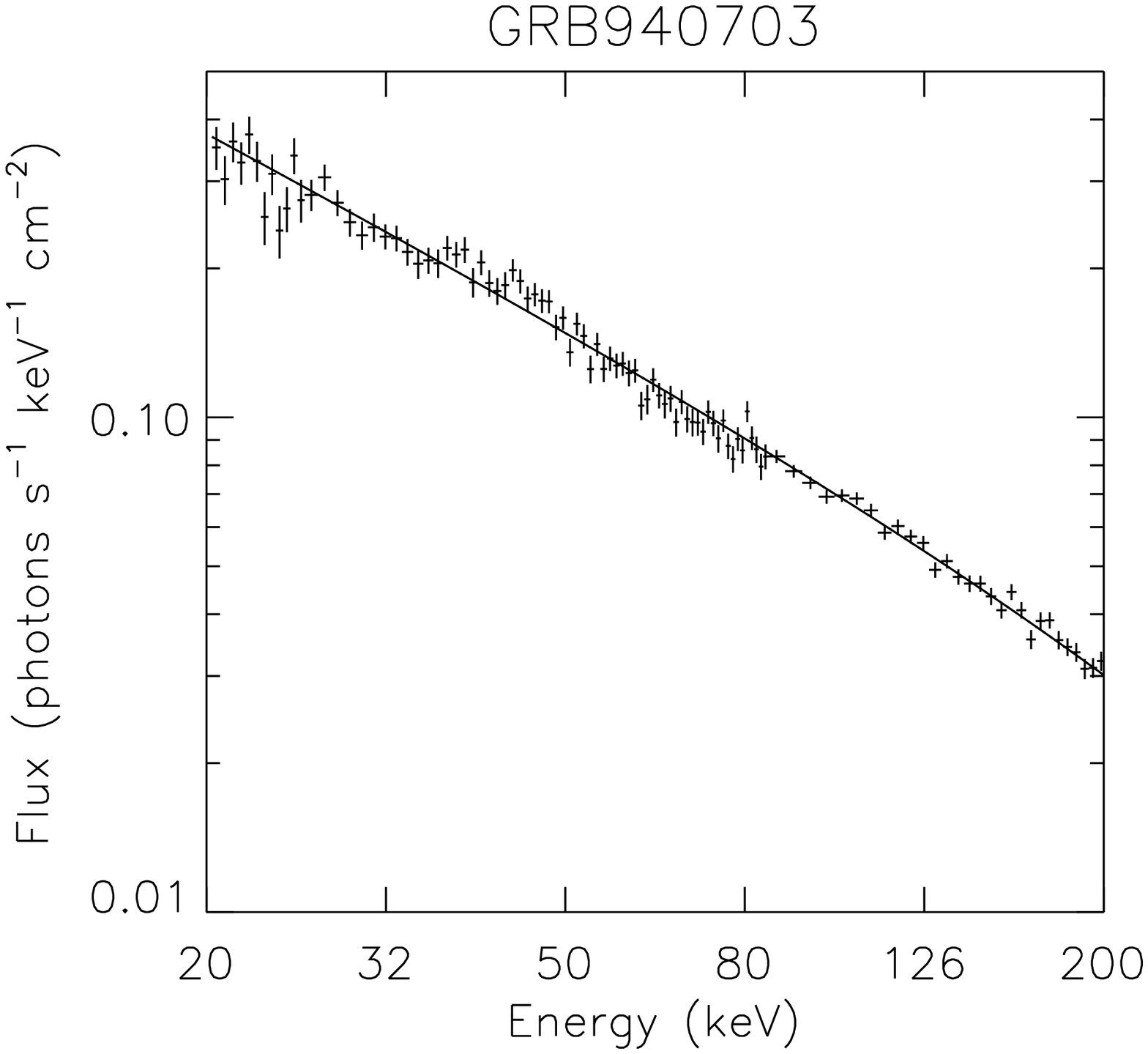,height=5.7cm,
bbllx=30bp,bblly=190bp,bburx=555bp,bbury=645bp,clip=}
\epsfig{file=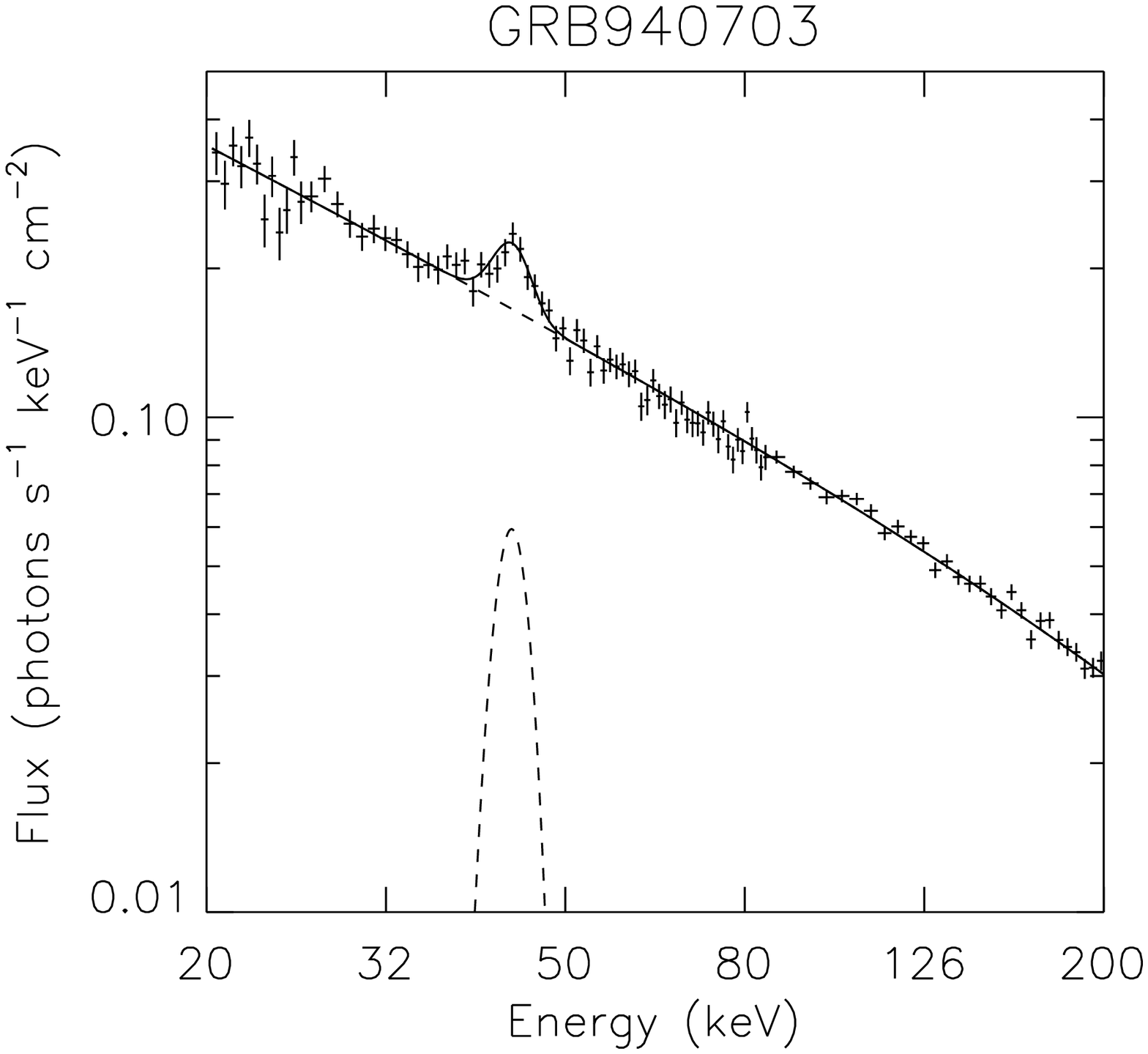,height=5.7cm,
bbllx=65bp,bblly=190bp,bburx=555bp,bbury=645bp,clip=}
}
}
\caption{Two fits to the same data of GRB~940703 are shown
(see Fig.~6 for the corresponding graphs of the count rate data).  
The curves show the
photon model and its components (dashed lines).  The
points in photon flux units are obtained by scaling the observed count
rate with the ratio of the model photon rate and model count rate.
It should be obvious that the points in photon flux units are
{\it model dependent} and should not be thought of as data.
}
\end{figure}

If we can never prove a fit to be `correct', how can one demonstrate the
existence of a spectral line?
The answer is two-fold: by viewing the problem as one of statistical
model comparison and by using scientific judgment.
The statistical approach used is model comparison:
a potential line is viewed as an additional term added to
the continuum model and we ask whether the fitting statistic (e.g.,
$\chi^2$) is sufficiently improved to convince us of the reality of the line.
Traditionally, the F-test has been used, but statistical theory shows
that the change in $\chi^2$,  $\Delta \chi^2$,
is better
(this test is also known as the Maximum Likelihood Ratio test 
(Freeman et al. 1999a)).   In practice, there is little different
between the two statistics (Band et al. 1997).
Bayesian analysis procedures have also been developed
(Graziani et al. 1993, Freeman et al. 1999a).
In principal, to demonstrate the existence of a line one must show that the
line is significant for any continuum model.
Scientific judgment can be used to limit the space of trial continuum
models to all `reasonable' models.

Some researchers believe that only continuum
models justified for the particular
spectrum should be used (Freeman et al. 1999a)---but
if a reasonable continuum model explains that data
without resort to a line, then we cannot be sure whether the line exists.
Consider two continuum models, `A' and a more complicated model `B'.
Further suppose that both models have acceptable $\chi^2$ values
when fitting a particular spectrum, and that adding a line creates
a significant $\chi^2$ improvement for model `A' but not for model `B'.
Despite the line being significant if model `A' is assumed, overall the
line must be considered insignificant because model `B' might be the
correct model.
Judgment enters the problem in deciding what the set of all reasonable
continuum models is.
Currently, the most popular continuum model is the four-parameter
`GRB' function of Band et al. (1993).   This function  fits
the available gamma-ray data well.
Most of the popular simpler
models can be represented as special cases of the GRB function;
depending on the SNR and energy range
of a spectrum, it may not be possible to demonstrate that the four-parameter
GRB function provides a statistically better fit compared to a simpler model.
It may be sufficient to use just the GRB function as the set of
all reasonable continuum models.
However, a recent analysis of data extending below 20 keV
indicates that in some cases the GRB  function may fail in the X-ray range
(Preece et al. 1996).

A final consideration is that recent experiments collect a large number 
of spectra from a large number of GRBs.   If all of these spectra 
and the additional spectra formed by summing them are
searched for lines, the possibility of a chance fluctuation mimicking a line
is increased.   This must be allowed for in calculating the overall
significance of a possible line feature.

The preceding discussion is based on the assumption that the instrument
works correctly, that the computer model of the detector is sufficiently
accurate, and that the errors in the observed spectrum arise solely
from Poisson fluctuations in the arrival of photons and their interaction
in the detector.    Systematic errors need to be limited by ground and
space based testing of the detectors.

If multiple detectors observed a burst with a possible line feature,
the feature needs to be tested by comparing the results from each detector.
If the feature is a statistical fluctuation, 
a matching fluctuation in a different detector is improbable.
Some types of systematics errors, e.g., hardware problems with one
detector or failures of the detector model as a function of source angle,
would also be revealed by discrepancies
between the data collected with different detectors.
After allowing for statistical fluctuations,
the data of the detectors must be consistent; ideally more than one detector
would detect the feature with statistical significance, thereby confirming
the presence of the line.

In summary, the steps to demonstrate the existence of a gamma-ray line are:
\begin{list}{$\bullet$}{\setlength{\itemsep}{0pt} \setlength{\parsep}{0pt}}
\item Deconvolve the spectra using the forward-folding technique,
\item Show the model fit and the data in the units of observed energy
loss (also known as count spectra), rather than in deconvolved photon flux
units,
\item Test the necessity of the line by comparing fits with and without
the line,
\item Test the significance of the line against all reasonable continuum models,
\item Quantify the model comparisons by using a statistic such as
$\Delta \chi^2$ or likelihood,
\item If a large number of spectra have been examined, consider the
increased probability of a chance fluctuation and appropriately
degrade the significance of a line candidate,
\item If several instruments or detectors observed an event, 
the datasets need to be analyzed for consistency.
\end{list}

\section{The First Era}

The work of the Mazets group created the field of GRB lines.
In 1981 they reported absorption lines between 30 and 70 keV in numerous bursts
in observations made with
the Konus experiments on the Venera 11 and Venera 12 spacecraft 
(Mazets et al. 1981).
Lines were reported to be a common characteristic of GRBs, with
low-energy lines observed in 43  of 143 GRBs (Mazets et al. 1982).
The Konus instruments consisted of six NaI(Tl) detectors, each 30 mm thick and
80 mm in diameter.
Lines were also reported in the Konus instruments of Venera 13 and 14,
which had similar detectors but improved electronics with better
temporal and spectral resolution (Mazets et al. 1983).

The lines appear as large deviations in a few channels from the assumed 
continuum model.
Most of the lines are absorption lines between 30 and 70 keV
with equivalent widths of 10 to 20 keV.
Only one low-energy feature appeared in emission.
Most of the lines are seen in only a portion of the burst, and that
portion is typically the beginning of the burst (Mazets et al. 1982).
Excepting a single 45 keV emission line (Mazets et al. 1981),
all of the emission lines are between 350 and 450 keV.
In several GRBs high-energy emission lines are present in the fits
to the data of both Venera 11 and 12 (Mazets et al. 1982).
These high-energy lines have intrinsic widths of a few hundred keV
and were interpreted as 
gravitationally redshifted annihilation radiation.

The analysis procedure was designed to minimize the computational effort:
spectra were deconvolved with standard templates and then iteratively
improved (Mazets et al. 1983).   Almost all of the spectra are depicted
in deconvolved photon units, making it difficult to judge the robustness of the
features and their significance if a different continuum model were to be
used.
The paper best describing the analysis procedure (Mazets et al. 1983) includes
a graph showing the observed count spectrum and the corresponding
deconvolved photon spectrum for an interval of GRB820406b which appears
to have an absorption feature at 45 keV.
At the time, GRBs were nearly always modeled with a simple
optically-thin thermal bremsstrahlung form,
photon flux $\propto E^{-1} \exp{-E/kT}$; this function was assumed
for most of the Konus spectra.
We now know that many GRB spectra have a high-energy power law,
the ``$\beta$-component'' of the Band GRB function, which cannot be
described by a function with an exponential cutoff, and that the
low-energy portion of the spectrum
(the $\alpha$ and $E_{\rm peak}$ parameters) undergoes
rapid spectral evolution.
These improvements in our knowledge of the
continuum function render interpretation of the Konus results difficult.

Efforts were made to validate the Konus results with the observations 
of other instruments.
Most of the comparisons focused on the high-energy emission lines.
Observations with the SIGNE experiments on the Venera spacecraft
were taken to show the presence of several lines in GRB~781104 and
GRB~781119 (Barat 1983).
Because of differing time intervals, the results were not directly
comparable with the lines reported by Mazets et al. (1981).
Data collected with the Gamma-Ray Spectrometer on the SMM spacecraft
for GRB~811231
included a spectrum collected over an interval overlapping two intervals
in which Konus data (Mazets et al. 1983) showed a high-energy emission line.
The data from GRS is consistent with a power law above 300 keV
(Nolan et al. 1981).

While there was some debate about the reality of the Konus lines,
the matter was viewed by most as settled when line observations with
the Ginga instrument were reported.
The Gamma-Ray Burst Detector for the Ginga satellite covered the energy
range 2 to 400 keV by using both a proportional counter (effective area
63 cm$^2$) and a NaI scintillator (1 cm thick, 60 cm$^2$ area)
(Murakami et al. 1989).
Harmonically spaced lines at 20 and 40 keV were seen
at high significance in GRB 880205 and at
lesser significance in GRB 870303 (Murakami et al. 1988,
Fenimore et al. 1988).
The harmonic spacing was seen at the time as powerful evidence
in support of cyclotron resonant scattering, implying an origin
on highly magnetized neutron stars.
The analysis of GRB 880205 was carefully done using the
forward-folding technique and using several continuum models,
including a very flexible continuum model, a power-law with
two breaks (Fenimore et al. 1988).
The F-test indicates a significance of $9 \times 10^{-6}$ for the pair of lines
in GRB 880205 (Wang et al. 1989).

Later analysis found lines at 26 and 47 keV
in GRB 890929.   Assuming harmonic spacing, a good fit was obtained
with centroids of 24 and 48 keV (Yoshida et al. 1992).
An additional interval with a 20 keV feature was found in GRB~870303
(Graziani et al. 1992).
With lines in three GRBs of 23 observed (Yoshida et al. 1992), Ginga also showed
low-energy lines to be a common feature of GRBs.

Two recent papers (Freeman et al. 1999a, Freeman et al. 1999b) have
very detailed analyses of the statistical significance of the lines 
observed with Ginga in
GRB~870303 and of their interpretation as cyclotron resonant scattering.

Lines were reported in the data of HEAO A-4 for three GRBs,
but only in the case
of GRB~780325 are the lines statistically significant (Hueter 1987).
The HEAO A-4 instrument included three types of NaI detectors to
cover the range 10 to 6200 keV.
GRB~780325 lasted $\approx 50$~s and consisted of two peaks: 
a 70 keV absorption feature was reported in the first peak and
a 50 keV absorption feature in the second.
The changes in $\chi^2$ were 14.0 and 16.5, respectively,
corresponding to chance probabilities of $9\times10^{-4}$
and $3 \times 10^{-4}$.
Most interestingly, the line is stated to be visible in the data of
both detectors which observed the burst in the energy range of the line,
however, only summed data is shown.
Unfortunately, some approximations were used that
cause the analysis to fall short of current standards:
a exponential continuum model was fit to the data below 200 keV
and a simplified detector model was used, created from the full detector
model by assuming an $E^{-2}$ power law  (Hueter 1987).
A figure of the 50 keV feature for a slightly different time interval
is available in Harding, Petrosian \& Teegarden (1986).

A pair of absorption lines were reported in GRB~890306, based on
data from Lilas (Barat 1993).
The Lilas detector was a NaI crystal 5.3 cm in diameter and 3 cm thick
covering the energy range 5 keV to 1 MeV.
The intense burst GRB~890306 lasted more than 70~s; the
line candidates appear in a spectrum accumulated over 68.5~s.
The fitting is done using the forward-folding approach and the data but
not the model are shown as an energy loss (count) spectrum.
A fit to the sky background is shown to allay concern about possible
systematics.
A very flexible spectrum was assumed, a power law with two breaks.
Adding two lines at 11 and 35 keV reduces $\chi^2$ from 132.2 to 43.2 for 26
degrees-of-freedom, with an implied chance probability of
$2 \times 10^{-13}$ by the F-test.
Possible concerns are the still-high value of $\chi^2$ and the possible impact
of the unknown location of the burst.

The HEAO and Lilas results have received less attention than they
deserve; both merit reanalysis using forward-folding fitting with a full
detector model and assuming several continuum models, including
the Band `GRB' function.

\section{BATSE Results} 

The addition of the Spectroscopy Detectors (SDs) 
to the BATSE instrument was motivated by the line results from the
Konus instruments.
There are eight SDs, each a 12.7 cm diameter by 7.6 cm thick NaI crystal.
The BATSE team expected to easily find numerous lines in GRBs.
Our first approach was to manually examine selected spectra from bright GRBs,
scanning for  possible features.   Spectra with possible features were fit
with continuum models and continuum-plus-line models to
evaluate the statistical significance of the feature.
No significant features were found by this technique (Palmer et al. 1994,
Band et al. 1996).

BATSE has several advantages for detecting lines:
\begin{list}{$\bullet$}{\setlength{\itemsep}{0pt} \setlength{\parsep}{0pt}}
\item Excellent resolution for NaI detectors,
\item Advanced electronics to handle large pulses and high counting rates,
\item Excellent temporal resolution to enable detection of lines on
many time\-scales,
\item The Large Area Detectors provide locations to aid analysis of the
spectral data from the Spectroscopy Detectors,
\item Multiple detectors to increase the chance of observing a line and
to allow consistency tests.
\end{list}
Simulations show that   BATSE should be able to detect
lines  like the 40 keV lines seen by Konus and Ginga 
with comparable sensitivity to Ginga (Band et al. 1995,
Freeman et al. 1993).
Ground and space-based tests demonstrate
that the detectors are working as expected (Paciesas et al. 1996).

Our failure to find lines was surprising.
To make sure that the manual search was not at fault, a 
comprehensive, automatic computer search procedure was developed
(Briggs et al. 1996).
The goal of the automatic search is to thoroughly search the spectra
collected from bright bursts so that no line could be accidentally missed.
Because we do not know a priori  when in a burst a line will occur
or for how long it will last, we search essentially all available timescales.
The individual  high-temporal resolution spectra from the SDs 
(the SHERB data) are all
searched, as are all consecutive pairs, triples, groups of five,  etc.,
up to the entire duration of the SHERB data.
The search is limited by choice to finding lines below 100 keV.
Since we also do not know a priori at what energies lines will occur,
we fit a continuum model to each spectrum, then continuum-plus-line models
with line centroids closely spaced over the available data up to 100 keV.
Because the well-calibrated data typically begins at 20 keV, the search
is insensitive to lines with centroids of 20 keV.

At this time 117 bright GRBs that were observed before 1996 May 31 have 
been examined.   An average of 2.1 Spectroscopy Detectors observed each
burst, producing a total of 10,942 SHERB records.
These records were combined into 120,700 spectra, many of which overlap
or have low signal-to-noise ratios (SNR).
We estimate the number of independent spectra with sufficient SNR
to enable the detection of a Ginga-like line as a few hundred to a several
thousand.

From the results of the automatic search, 17 candidates with $\chi^2$
changes from adding a line above $\Delta \chi^2 =20$ were identified.
Manual fitting of the data leaves 11 candidates with $\Delta \chi^2$ values
ranging from the threshold of 20 to 50.8.
Since we assumed lines with narrow intrinsic widths compared to the detector
resolution, the line fits involve only two parameters, centroid and amplitude.
There are about 5 independent energy resolution elements between the detector
threshold and 100 keV.
This implies that the candidates have fluctuations probabilities in a single
spectrum of $2 \times 10^{-4}$ (for $\Delta \chi^2 =20$)
to $5 \times 10^{-11}$ (for $\Delta \chi^2 = 50.8$).
With at most several thousand bright, independent spectra, the ensemble
probability of the most significant event is $\sim 10^{-7}$.

The candidate spectral feature in GRB~940703 (BATSE trigger 3057)
(Briggs et al. 1996)
has the following properties:
1) the significance is high, $\Delta \chi^2 = 31.3$, corresponding
to a fluctuation probability in a single spectrum of 
$8 \times 10^{-7}$,
2) the interval in which the line is most significant 
is close to the entire flux of the GRB,,
3) the feature is an emission line,
4) the centroid is 44 keV.
The appearance of the feature is shown in Fig.~6.
While there are exceptions, these properties are quite typical of the
other candidate features.


\begin{figure}[tb!]
\centerline{
\mbox{
\epsfig{file=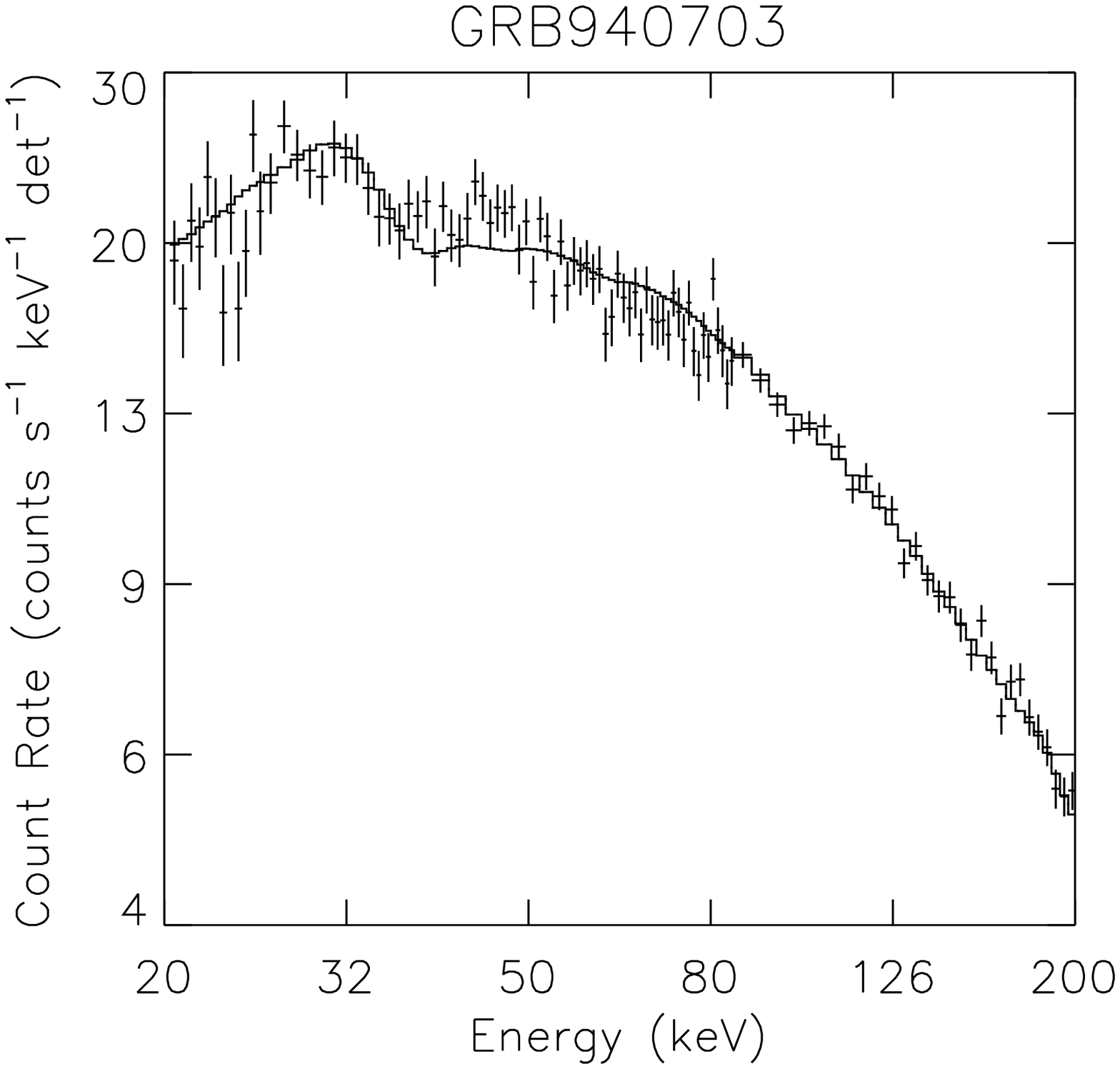,height=5.7cm,
bbllx=50bp,bblly=195bp,bburx=555bp,bbury=645bp,clip=}
\epsfig{file=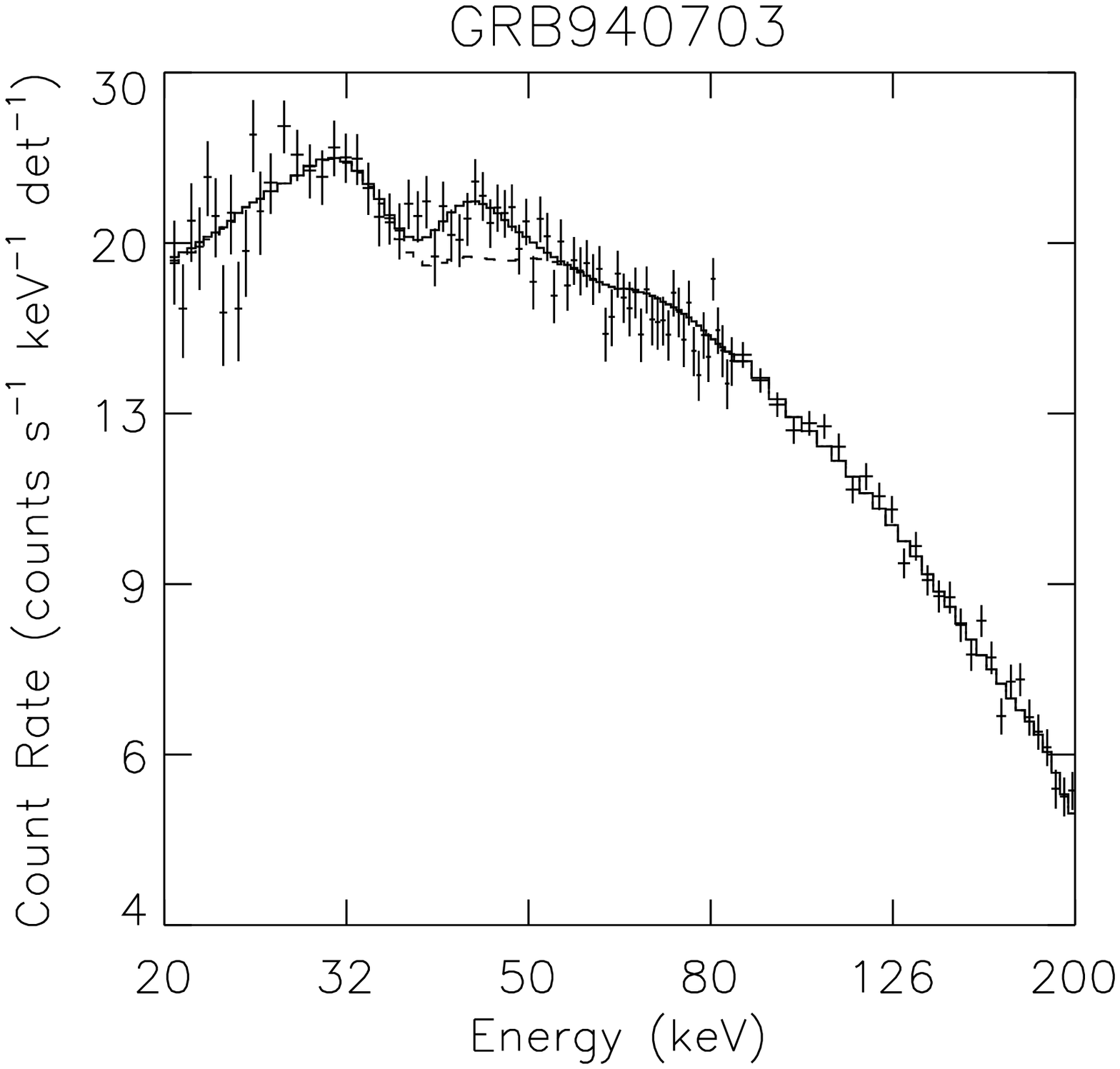,height=5.7cm,
bbllx=85bp,bblly=195bp,bburx=555bp,bbury=645bp,clip=}
}
}
\caption{
Two fits to the data of GRB 940703 from SD~5 for a 38~s interval.
Left figure: continuum-only fit using the Band `GRB' function,
Right figure: continuum-plus-line fit adding a narrow Gaussian line.
The data and model are shown as energy loss or count rate spectra;
the corresponding deconvolved spectra are shown in Fig.~5.
The data are shown as points with the vertical error bars showing the
uncertainties due to Poisson fluctuations and the background model.
The histograms depict the model: the solid line is the total model,
while the dashed line show the continuum-only contribution.
The feature is assumed to be narrow---the width in count space originates
from the spectral resolution of the detector.
}
\end{figure}

The final stage of the analysis is the comparison of the observations
collected with all of the BATSE Spectroscopy Detectors.
In the case of GRB~940703, this analysis cannot be done because 
none of the other SDs have useful data: either the detector gains were
inappropriate or the burst was viewed at too large of an angle.
Most of the other candidates have useful data from more than one SD.
The comparison analysis is complex because we are not comparing the
data from one detector to a known model, but rather comparing the data
of two or more detectors, not knowing the photon flux that created the
data.   For example, it is quite possible to have line flux values obtained
from different detectors that disagree because of upward and downward
fluctuations.
The consistency or inconsistency of
differing fit results  can only be determined with simulations.

Our procedure has been to jointly fit the data from all the SDs with
good angles and gains, and to assume that the photon model parameters
so obtained are a good representation of reality.
We then use this photon model and the detector models to create simulated
count data incorporating Poisson fluctuations.
Many simulations are created and fit in order to determine the range
of results expected.
The actual data are compared to expected range of results to determine
whether the data from all of the detectors is consistent with a common
origin.
Because the data of at least one detector indicate a statistically significant
line, if the data from all of the detectors are consistent with the common
photon spectrum, then there is strong evidence for a spectral feature.
Conversely, if the data from two or more detectors is inconsistent,
the interpretation becomes uncertain---at least one of the detectors
is suspect, either because of hardware problems or an inadequate
detector model.

The best case is that of GRB~941017 (trigger 3245).
Good data is available from both SD~0 and SD~5;
a feature was found by the automatic search in the data of SD~0 and
the data of SD~5 appear consistent (Briggs et al. 1998).

However, there are other cases in which the consistency is poor.
In GRB 930916 (trigger 2533), the data of SD~2 (Fig.~7) contain a 
highly significant line feature (Briggs et al. 1999).
Useful data is also available from SD~7 (Fig.~8), but this data contains
no indication of a line.  Furthermore, adding a line with the
centroid and strength obtained from the fit to the data of SD~2 actually
increases the $\chi^2$ of the fit to the data of SD~7.
The consistency between these two datasets has been analyzed via simulations
(Briggs et al. 1999): assuming the model parameters obtained from a joint
fit to the data of SD~2 and~7, simulated datasets are created for both
detectors.   These simulated datasets are then fit to determine the
expected range of line strengths.
In only 9\% of the simulations of SD~2 is a $\Delta \chi^2$ value above
23.1 obtained,
indicating that the observed line signal is somewhat stronger than expected.
Conversely, in the simulations of SD~7 a $\Delta \chi^2$ value below
0.1 is obtained
in only 2\% of the cases, indicating that observed line signal is weaker
than expected.
These two results are at best marginally consistent.

\begin{figure}[tb!]
\centerline{
\mbox{
\epsfig{file=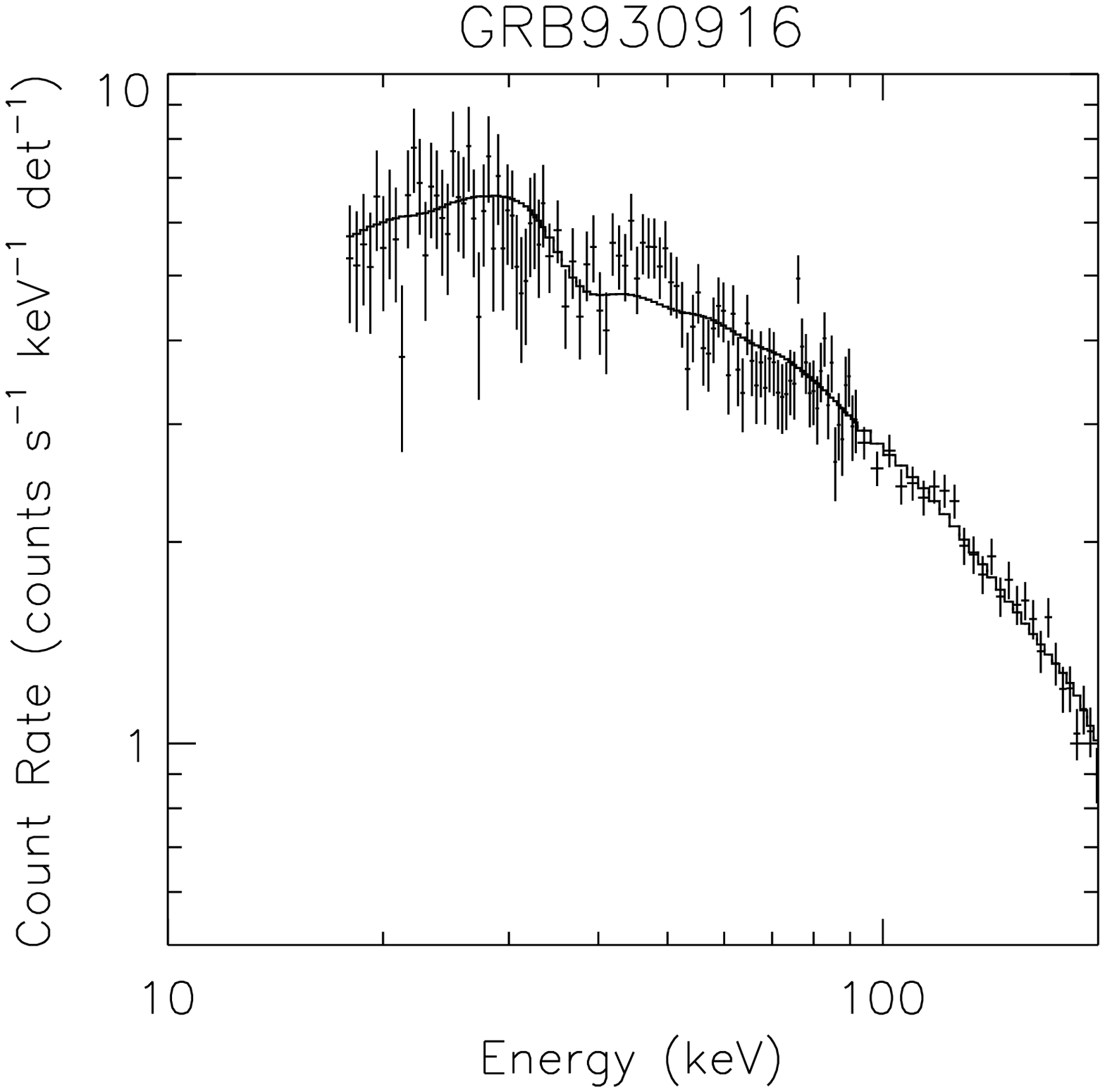,height=5.7cm,
bbllx=45bp,bblly=195bp,bburx=555bp,bbury=650bp,clip=}
\epsfig{file=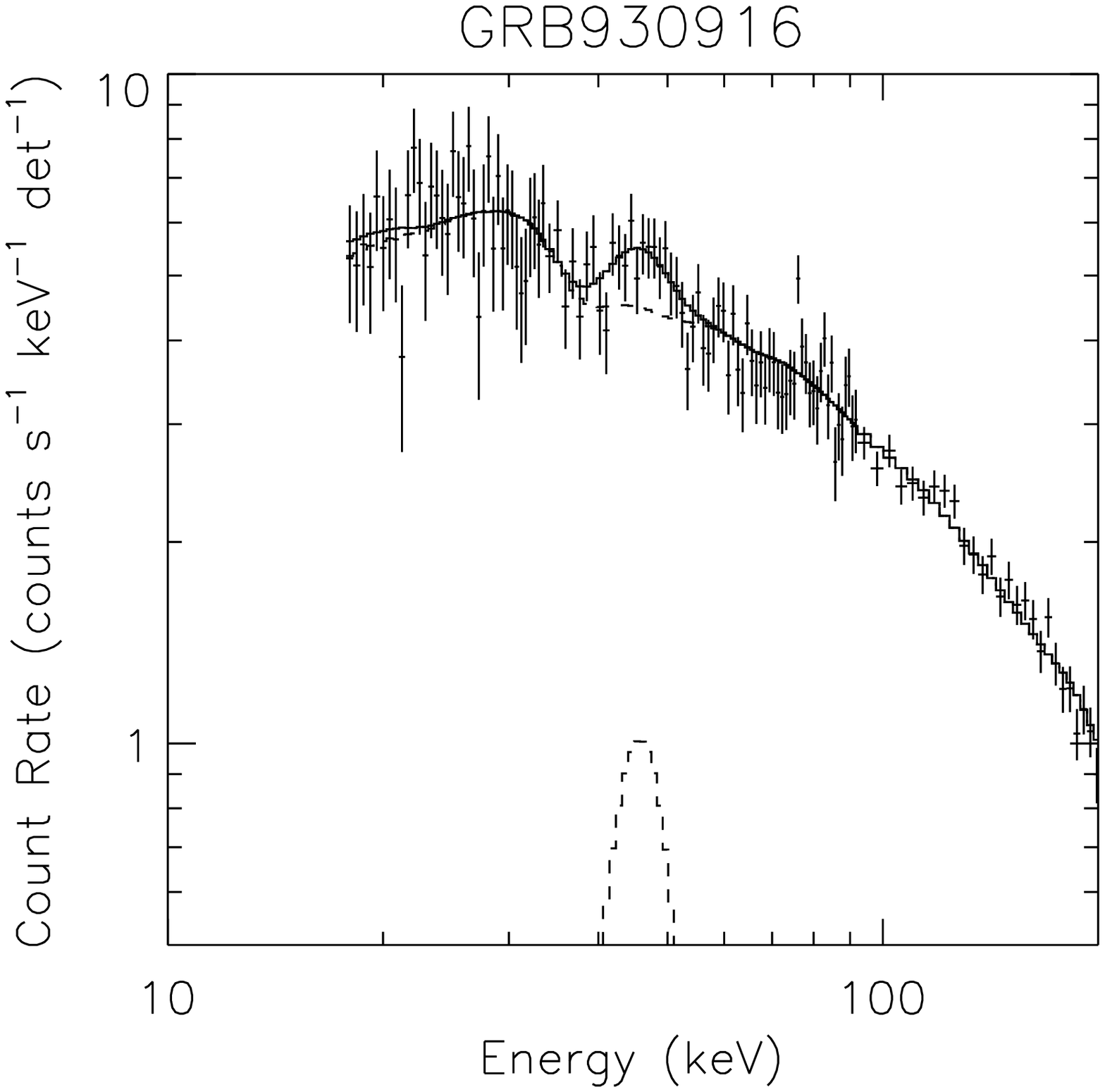,height=5.7cm,
bbllx=85bp,bblly=195bp,bburx=555bp,bbury=650bp,clip=}
}
}
\caption{Data from SD~2 for a 61~s interval of GRB 930916.
Left panel: a continuum-only fit using the Band GRB function.
There is an obvious cluster of data points above the model from
41 to 51 keV.  Right panel: A narrow Gaussian line is added to the
model.   The emission line with a centroid of 45 keV reduces $\chi^2$
by 23.1, corresponding to a chance probability of $5 \times 10^{-5}$.
}
\end{figure}

If GRB~930916 were the only such case, the agreement might be considered
acceptable.  Unfortunately, this level of disagreement occurs for
several other candidates.   The poor agreement between the detectors leaves
us uncertain of which to believe: the data which seems to indicate the
existence of a line of strength $S$,
or the data which seems to show the absence of a line of strength $S$.
Until this discrepancy is resolved, the meaning of the BATSE results
will remain unclear. 

\begin{figure}[tb!]
\centerline{
\mbox{
\epsfig{file=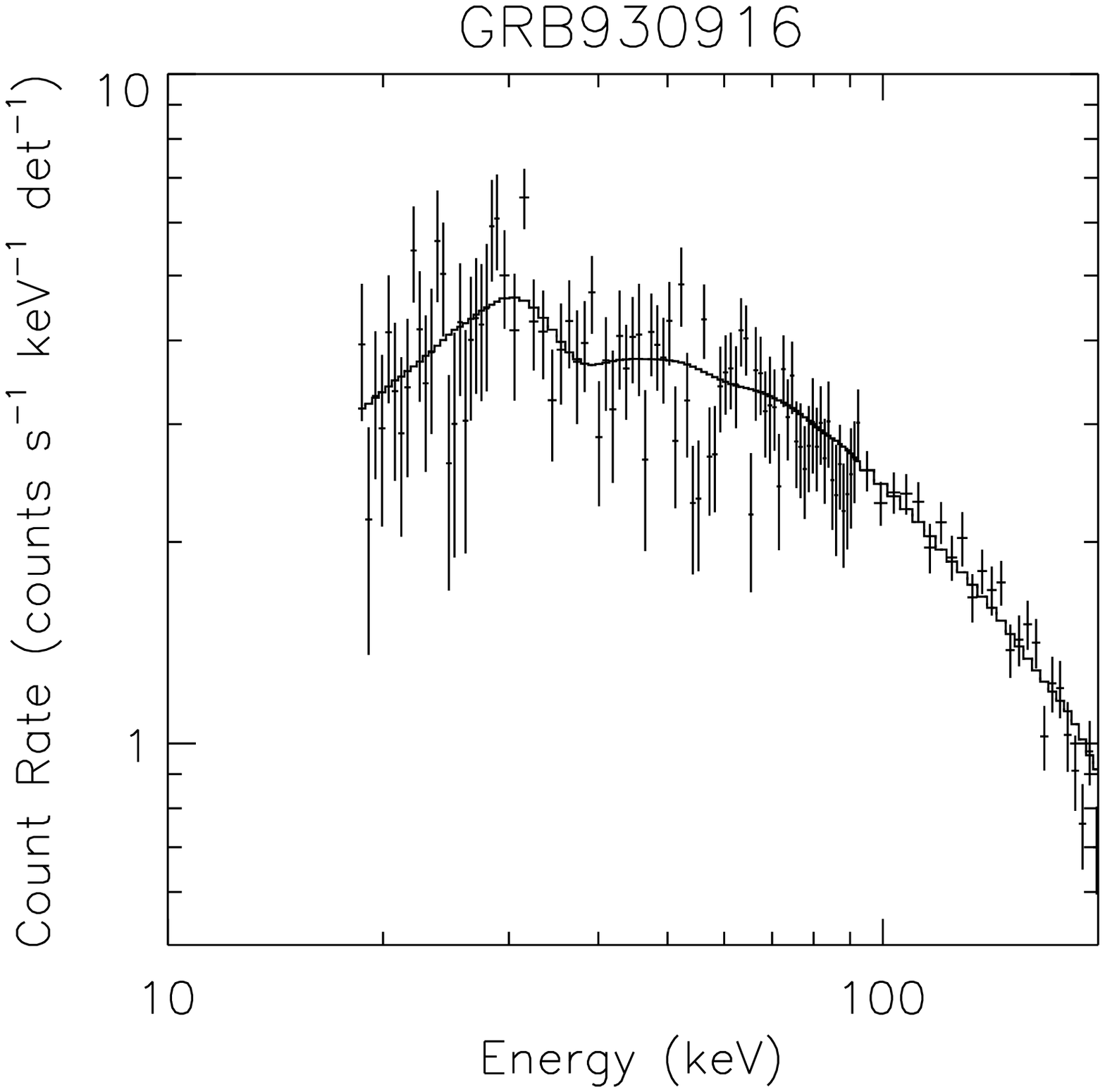,height=5.7cm,
bbllx=45bp,bblly=195bp,bburx=555bp,bbury=650bp,clip=}
\epsfig{file=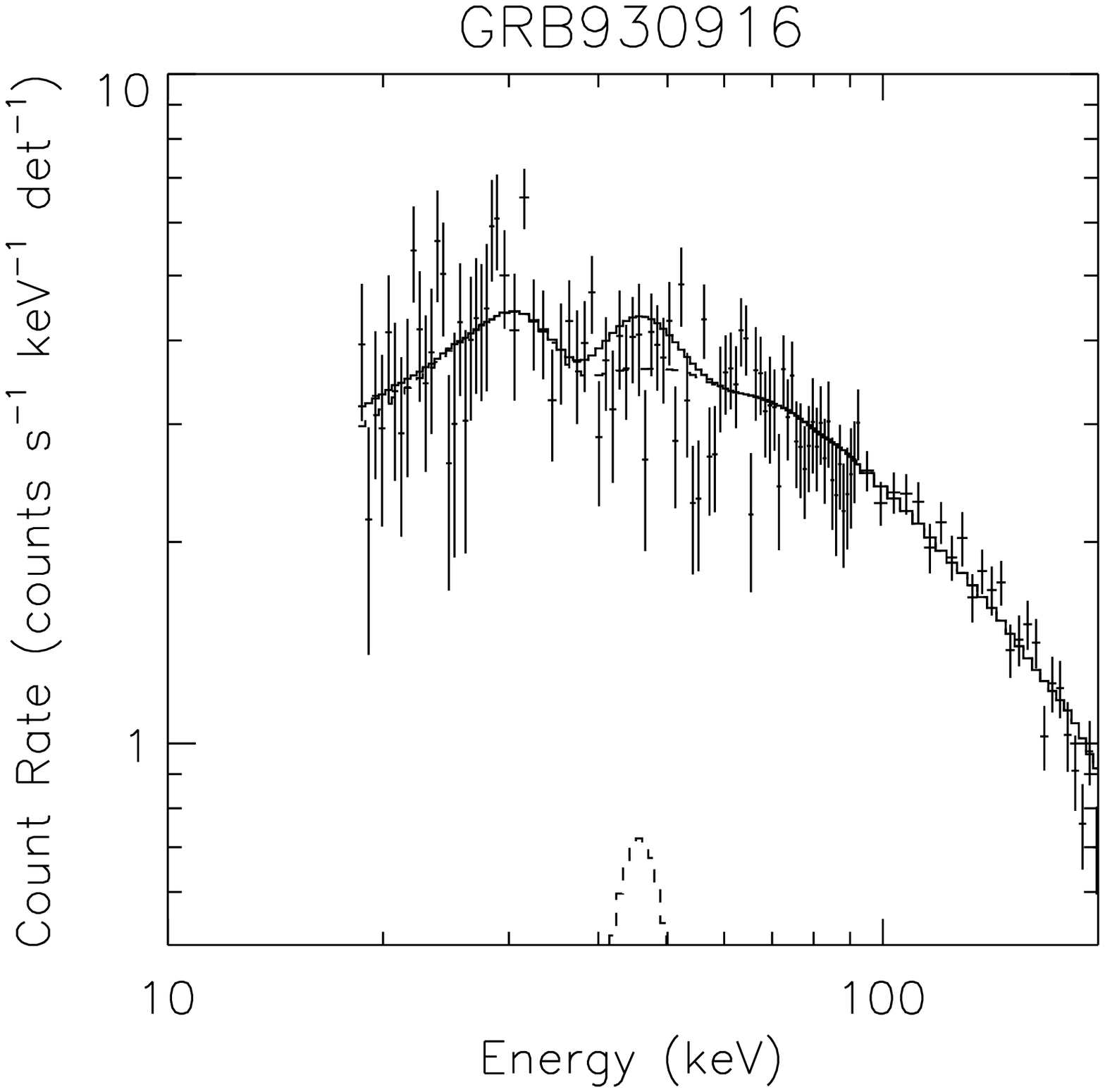,height=5.7cm,
bbllx=85bp,bblly=195bp,bburx=555bp,bbury=650bp,clip=}
}
}
\caption{
The data of SD~7 for the same interval of GRB 930916.
Left panel: Continuum-only fit.   There is no cluster of points deviating
from the continuum model; adding a line results in no improvement in
$\chi^2$ value.
Right panel: a line has been added to the model, 
with intensity and centroid fixed at the values 
inferred from the data of SD~2.
This fit increases $\chi^2$ by 9.7.
}
\end{figure}

\section{Other Recent Results}

There are two GRB instruments on the WIND spacecraft, which has a low and
stable background because of its location in interplanetary space.
Konus-W consists of two oppositely oriented NaI detectors which are
very similar to the BATSE SDs (Aptekar et al. 1998).
Since November 1994 at least 20 GRBs have been identified as containing
possible line candidates.
Golentskii et al. (1998) show energy loss spectra and deconvolved
photon spectra for three possible absorption features and one possible
emission feature.
Final statistical analysis is in progress.

The WIND spacecraft also has a gamma-ray burst detector using a
cooled germanium crystal, the Transient Gamma-Ray Spectrometer (TGRS).
This detector is characterized by a 
better spectral resolution than
scintillators but a rather small effective area, 36 cm$^2$.
Comparing to the BATSE Spectroscopy Detectors, the better resolution but 
lesser area of TGRS causes TGRS to be more sensitive to lines narrower
than the resolution of the BATSE detectors but less sensitive to lines
of width comparable to the resolution of the BATSE detectors
(Kurczynski et al. 1999).
A search of 36 bright events, not all of which are GRBs, found no
significant lines with centroids from about 40 keV to a few hundred keV
(Kurczynski et al. 1999).

In the 1980's there were many papers on the physics of line formation
via cyclotron resonant scattering on nearby highly magnetized neutron
stars.
With the evidence for a cosmological distance scale for most or all GRBs,
the picture of the physical conditions has changed: the outflow must
be highly relativistic.
Several recent papers treat the subject of line formation in
relativistic outflows, either in sources in the galactic halo,
a possible origin for some GRBs (Isenberg et al. 1998), or in a cosmological
fireball (Hailey, Harrison \& Mori 1999).
The later paper attempts to explain Ginga-like lines as complex ionization
spectra emitted by high density material entrained in a relativistic fireball
at cosmological distances, observationally smeared by the spectral
resolution of the detectors.
Much more work could be done on the possibility of line formation
in cosmological sources; one fundamental problem is that a range
of Lorentz factors should preclude the formation of any narrow feature.

\section{Conclusions}

The observational status of GRB lines is mixed: in the 1980's
several instruments reported low-energy absorption
lines to be present in a large fraction
of all GRBs, while there has been a dearth of recent detections.
The BATSE database contains $\sim$10 highly significant line features,
but these are low-energy {\it emission} features.
More disturbingly, in several cases the data from the several BATSE detectors
that viewed an event appear inconsistent.
In hindsight, it appears that insufficient attention was paid to
possible systematics, such as the correct continuum model, detector
performance and detector modeling.
Comparisons between the data of the BATSE Spectroscopy Detectors are
a start on testing these problems.
Further progress will be made by comparing data from several instruments,
e.g., BATSE and Konus-W, and by observations with future instruments.

In the 1980's GRBs lines were seen as important or even conclusive evidence
that GRBs originate from nearby (100 pc scale) highly-magnetized neutron
stars.
The BATSE evidence of the combination of isotropy with a deficiency of
bright bursts (inhomogeneity) cast strong doubt on this picture.
With the measurement of redshifts in some GRB afterglows, the 
paradigm has shifted to that of GRBs originating at cosmological distances.
This has created an excessive biases against GRBs lines---we should
consider the spectral observations on their own merit and remember the
possibilities that there might exist a galactic subclass of GRBs
or that lines might form in
cosmological GRBs via some as yet unthought of process.

\end{document}